# An Analytic Linear Accelerator Source Model for Monte Carlo dose calculations. II. Model Utilization in a GPU-based Monte Carlo Package and Automatic Source Commissioning

Zhen Tian[1], Michael Folkerts[1], Yongbao Li[1,2], Feng Shi[1], Steve B. Jiang[1], Xun Jia[1]

[1]Department of Radiation Oncology, University of Texas Southwestern Medical Center, Dallas, TX 75390, USA

[2]Department of Engineering Physics, Tsinghua University, Beijing, 100084, China

Emails: zhen.tian@utsouthwestern.edu, steve.jiang@utsouthwestern.edu, xun.jia@utsouthwestern.edu

We recently built an analytical source model for GPU-based MC dose engine. In this paper, we will present a sampling strategy to efficiently utilize this source model in GPU-based dose calculation. Our source model was based on a concept of phase-space-ring (PSR). The ring structure makes it effective to account for beam rotational symmetry, but not suitable for dose calculations due to the presence of rectangular jaw areas. Hence, we developed a method to convert PSR source model to its phase-space let (PSL) representation. In dose calculation process, different types of sub-sources were separately sampled. GPU kernel of source sampling and kernel of particle transport were iterated. This ensured that the particles being sampled and transported simultaneously are always of the same type and close in energy in order to alleviate GPU thread divergence. The second purpose of this paper was to present an automatic commissioning approach to automatically adjust the model to achieve a good representation of a clinical linear accelerator (linac). Weighting factors were introduced to adjust relative weights of PSRs. Determining these factors was realized by solving a quadratic minimization problem with a non-negativity constraint. We have tested the efficiency gain of our analytical source model over a previous source model using PSL files. It was found that the efficiency was improved by by 1.70 ~ 4.41 times in phantom and real patient cases, mainly due to the avoidance of long data reading and CPU-to-GPU data transferring. The automatic commissioning problem can be solved in ~20 sec. Its efficacy was tested by comparing the doses computed using the commissioned model, the uncommissioned one, with measurements in different open fields in a water phantom under a clinical Varian Truebeam 6MV beam. For the depth dose curves, the average distance-to-agreement (DTA) was improved from 0.04~0.28 cm to 0.04~0.12 cm for build-up region and the root-mean-square (RMS) dose difference after build-up region was reduced from 0.32%~0.67% to 0.21%~0.48%. For the lateral dose profiles, RMS difference was reduced from 0.31%~2.0% to 0.06%~0.78% at inner beam region and from 0.20%~1.25% to 0.10%~0.51% at outer beam region.

## 1. Introduction

The accuracy of Monte Carlo (MC) dose calculations (Ma *et al.*, 1999; Rogers, 2002, 2006; Keall *et al.*, 2000) depends on both the accuracy of particle transport simulations within the patient and that of the linac beam modeling. Recently, there has been a burst of research aiming at developing high performance MC simulation packages on a graphics processing unit platform (Jia *et al.*, 2010; Jia *et al.*, 2011; Hissoiny *et al.*, 2011; Jahnke *et al.*, 2012). While it has been demonstrated that particle transport simulations can be achieved in an accurate and efficient fashion, linac beam modeling for GPU-based MC simulations has been rarely reported.

A phase-space file based source model has been developed for a GPU-based dose engine gDPM (Townson *et al.*, 2013). The concept of phase-spacelet in this model enabled an automatic model commissioning method that finely tunes the model for an accurate representation of a linac beam (Tian *et al.*, 2014). However, it was found that this is not a favorable approach due to the relatively long time for data loading, transfer and processing. In contrast, conventional analytical source model is expected to be more preferred for GPU-based dose engines, in that long data transfer time can be avoided and it is quite efficient to sample particles on the fly. Over the last few decades, analytical source models have been extensively studied (Ma *et al.*, 1997; Ma, 1998; von Wittenau *et al.*, 1999; Deng *et al.*, 2000; Fix *et al.*, 2004; Verhaegen and Seuntjens, 2003; Davidson *et al.*, 2008). These models, however, are developed for the conventional CPU platform. While the model itself is applicable to the GPU context, some aspects of the model require special attention considering the parallel processing scheme in GPU, e.g. how to efficiently sample particles from the model.

We have recently developed an analytical source model specifically for GPU-based dose engine. In a series of two papers, we have presented in the first paper our model and a method to derive model parameters based on a phase-space file (Tian *et al.*, 2015a). In this current paper, we will address two other issues regarding the clinical applications of the model.

The first issue is how to efficiently sample a particle from our model on the fly of the GPU-based dose calculations. In fact, our source model consisted of a set of phase-space-rings (PSRs). Each PSR represented a group of particles that were from the same sub-source (primary photon, secondary photon, or electron), resided in a narrow ring in the phase-space plane, and were in a certain energy range. Particle probability density for each PSR was parameterized. The introduction of this PSR concept allowed us to use the rotational symmetry of a beam to help reduce degrees of freedom of the model. However, it was noted that direct use of this PSR model in dose calculations was not preferred due to three reasons. 1) For a treatment plan with a defined rectangular beam area, it was not necessarily to sample particles from all the PSRs in the model, as those particles passing through the jaw opening and hence contributing to the dose were only from a few PSRs. For instance, in the case shown in Fig. 1(a), the most outer ring is unnecessary in dose calculation. 2) Even we identified those PSRs sufficiently to cover the jaw opening area, sampling particles from these PSRs would unnecessarily involve those particles in regions that are far away from jaw opening, as shown in the shaded area in Fig. 1(a), reducing the overall efficiency. This would be particularly a problem for non-square and/or off-axis jaw settings in Fig. 1(b). 3) GPU adopts a single-instruction multiple data (SIMD) scheme (Jia *et al.*, 2014) to execute codes. It is hence preferred to design a sampling approach to coordinate the sampling processes among GPU threads in order to maximally comply with the SIMD scheme. In this paper, we will design a simulation workflow that enabled integration of this source model in a GPU-based dose engine and efficient dose calculations.

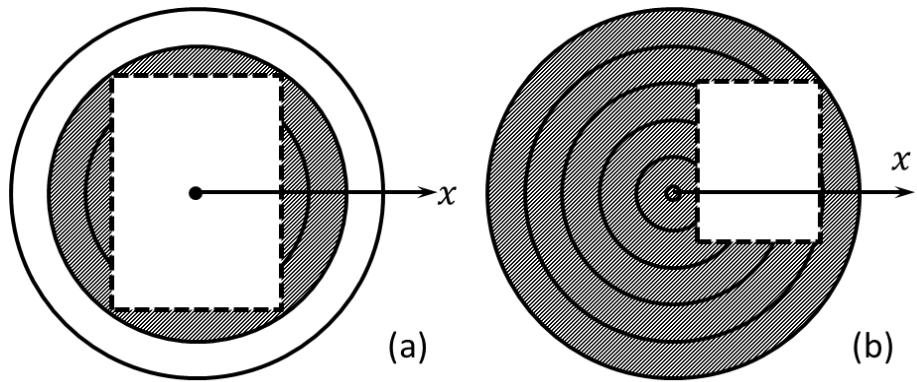

**Figure 1.** (a) When performing dose calculations using a PSR-based source model, particles in the shaded area will be rejected. The rectangular box corresponding to the jaw buffer determined by the jaw opening in a plan. (b) The sampling efficiency becomes much reduced for an off-axis and small field.

The second issue we would like to address in this paper is automatic model commissioning. We have previously presented a method to derive parameters in our model from a phase-space file. Yet it is expected that finely tuning the model is still necessary in clinical applications to account for the difference between the beam represented by the phase-space file and the actual beam used in the clinic. Manually adjusting those parameters in the model is impractical. An automatic beam commissioning approach is therefore desired. Hence, we will also propose in this paper an automatic commissioning method which solves this problem via an optimization approach.

**2. Methods and Materials**

*2.1 Model utilization*

While the PSR concept is beneficial to exploit the beam's rotational symmetry for model construction, as stated earlier, using it for dose calculations is less optimal due to the rectangular jaw settings, as illustrated in Fig. 1. In this regard, the original phase-space-let (PSL) concept (Townson *et al.*, 2013) is more suitable. Hence, we propose to first convert a PSR-based model to its corresponding PSL representation, and the latter is used for dose calculation. Note that this conversion step only needs to be done once, after the PSR-based model parameters is determined.

*2.1.1 Convert PSR model into its PSL representation*

Apart from a sub-source index $stype$ and energy index $Ebin$, a PSL is associated with two location indices $Xbin$ and $Ybin$. Since the conversion is only performed on location domain, the rest of derivations in this subsection are independent of energy bin and sub-source. These two indices are hence omitted to simplify the notation. To derive a PSL representation for a PSR model, there are two factors to consider, namely the weighting factor for each PSL and the particle direction distribution within it. In the continuum limit, let us denote the probability density for a particle with a radius $r$ as $p^{PSR}(r)$ and the probability density for a particle at a location $(x, y)$ as $p^{PSL}(x, y)$. Because of the rotational symmetry assumption, integration in a ring area with a radius of $[r, r + \Delta r]$ yields

$$\int_r^{r+\Delta r} p^{PSR}(r)\, dr = \iint p^{PSL}(x,y) dx dy |_{r^2 \leq x^2+y^2 \leq (r+\Delta r)^2} \qquad (1)$$
$$= \int_0^{2\pi} d\theta \int_r^{r+\Delta r} p^{PSL}(r\cos\theta, r\sin\theta) r dr$$
$$= 2\pi \int_r^{r+\Delta r} p^{PSL}(r\cos\theta, r\sin\theta) r dr \,.$$

This implies

$$p^{PSL}(x,y) = \frac{1}{2\pi r} p^{PSR}(r)|_{r=\sqrt{x^2+y^2}}. \qquad (2)$$

Hence, in the discrete form, the relative weighting factor $W_{Xbin,Ybin}$ can be converted from $W_{Rbin}$ as following. 1) Calculate the radius $r = \sqrt{x^2 + y^2}$ for this PSL, where the coordinate $(x, y)$ are understood as the coordinate for the PSL center. 2) Find out the weighting factor corresponding to this radius in the list of PSR weighting factors $W_{Rbin}$. Note that the radius $r$ corresponds to a PSL may fall between two successive PSR ring radii. We performed linear interpolation along the radial direction to handle this issue. 3) Divide the resulting weight by $2\pi r$, yielding $W_{Xbin,Ybin}$. 4) Normalize the weighting factors such that $\sum W_{Xbin,Ybin} = 1$, where the summation is over all the PSLs with different locations, energies, and sub-sources.

The particle direction part is relatively easy to handle. Due to the beam rotational symmetry assumption, our PSR-based source model defined a local particle direction coordinate at each point on the phase-space plane, such that the direction component $u$ is defined along the radial direction, and the component $v$ is in the phase-space plane but perpendicular to $u$. Hence, at each PSL the direction probability density is identical to the probability density of the PSR that this PSL center belongs to. This fact is sufficient for us to use the established directional distribution of PSRs to sample particle directions for each specific PSL, as will be shown in Sec. 2.1.3.

*2.1.2 Determine sample areas*

In a dose calculation process for a clinical treatment plan, not all particles on the phase-space plane can go through a given jaw opening. This fact motivates us to use the PSL representation. Under this representation, it is convenient to selectively sample particles only within a rectangular area that includes the actual jaw opening area and a buffer region surrounding it (Townson *et al.*, 2013). The purpose of including the buffer region is to account for scattered particles that are outside the jaw opening area on the phase-space plane but still can reach the patient. This is much more efficient than sampling all particles at the phase-space plane and rejecting them later, especially for cases with a small jaw opening. The union of the actual jaw opening and the buffer region is referred as "sampling region" hereon.

Different from our previous approach (Townson *et al.*, 2013), we defined the sampling area for primary and scatter sub-sources respectively, since primary and scattered components are separately considered in our source model. Specifically, for primary photons the sampling area is an area of jaw opening scaled back to the phase-space plane. This is illustrated in Fig. 2. Furthermore, some PSLs may be partially inside this scaled jaw area. We further enlarge the sampling area to cover all of those partially covered PSLs. As for the sampling area for the scattered sub-sources, we first compute the beam center location at the iso-center plane. We then back project the upper surfaces of the jaw opening area from this point to the phase-space plane. The upper surface of the jaws will also be used for scatter particle rejection in dose calculations, which has been shown to yield acceptable accuracy in our previous studies (Townson *et al.*, 2013). Since the two pairs of jaws are located at different *z* levels, we separately scale them and the sampling area is determined by the intersection between the sampling area for the x and the y directions. It was further enlarged to cover any partially involved PSLs.

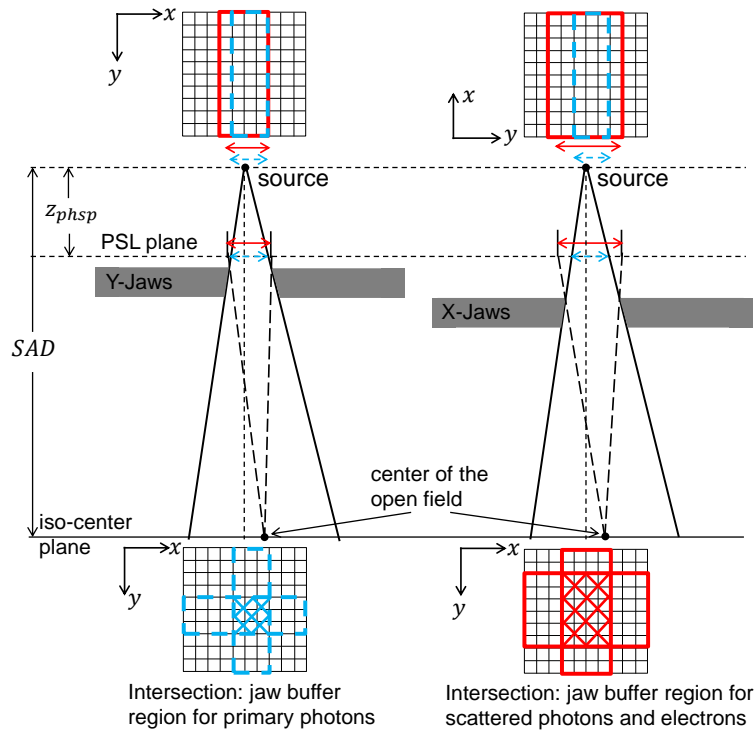

**Figure 3.** Illustration of how we determine sampling areas for primary photons, scattered photons and electrons, respectively. The sampling region for primary photons is determined directly by the equivalent jaw position on the PSL plane, while the region for scattered photons and electrons is determined by finding the center of the open field on iso-center plane and back-projecting from the center through jaw openings to the PSL plane. A PSL is only considered in subsequent dose calculation if any portion of its area is within the sampling region. The final PSLs sampled for dose calculation are the ones within the intersection area of those exposed by both X-jaws and Y-jaws.

*2.1.3 Main sampling strategy*

With the sampling area determined, we can then proceed to sample source particles and transport them for MC dose calculations. In a conventional CPU-based dose calculation case, we sequentially sample each source particle, which can be from one of the three sub-sources. However, for GPU-based dose calculations, since a number of threads sample source particles simultaneously and then transport them, it is desirable to handle different types of sub-sources separately. Otherwise different GPU threads would perform sampling particles for different sub-source types according to their own probability density model and transporting them without separating photons and electrons, yielding the thread divergence problem. With this in mind, our GPU-based dose calculation code has the following procedure.

S1. *Determine sampling areas.* Determine the sampling areas for each beam based on its jaw setting, according to the procedure stated in Sec 2.1.2. Then Repeat steps S2-5 for each beam in a treatment plan.

S2. *Determine number of particles for PSLs.* For those PSLs that are within the sampling area for the current beam, determine the number of particles $n_{stype,Ebin,Xbin,Ybin}$ for them based a user specified total number of particles $N$ to be sampled for the treatment plan. This is achieved by

$$n_{stype,Ebin,Xbin,Ybin} = \frac{N \times W_{stype,Ebin,Xbin,Ybin}}{\sum_{ibeam}\sum_{stype,Ebin,(Xbin,Ybin)\in A_{stype}} W_{stype,Ebin,Xbin,Ybin}}, \quad (3)$$

where $A_{stype}$ denotes the sampling area determined previously for each type of subsources, and the summation in the denominator is over all the sampling areas of all the beams.

S3. *Sample primary photon sub-source PSLs.* Sample in each primary photon PSL a given number of particles determined in Eq. (3) on GPU. Within each PSL, the photon location is uniformly sampled in its corresponding area and the photon energy is uniformly sampled in its corresponding energy range. Recall the probability density model of particle direction for primary photon PSRs (Tian *et al.*, 2015a),

$$p_{pp,i}(u,v|r,E) = \int \delta(u-u_0)\delta(v-v_0)\frac{1}{2\pi\sigma_s^2}e^{-\frac{x_s^2+y_s^2}{2\sigma_s^2}}dx_s dy_s, \quad (4)$$

$$u_0 = \frac{r-x_s}{\sqrt{(r-x_s)^2+y_s^2+z_{phsp}^2}}, \quad v_0 = \frac{-y_s}{\sqrt{(r-x_s)^2+y_s^2+z_{phsp}^2}}. \quad (5)$$

Once the photon location is determined, we first sample a source location $(x_s, y_s)$ within the beam spot from a 2D Gaussian distribution in Eq.(4). The model parameter $\sigma_S$ was stored on GPU's constant memory for fast access. The photon direction was then determined by Eq. (5). Note that the coordinate $(u,v)$ in our model is defined locally at each $(x,y)$ coordinate, such that $u$ is along the radial direction. A rotation of the particle direction coordinate is needed by an angle $\theta = \mathrm{atan}\, y/x$ to yield the coordinate for particle transport, namely $(u,v)$ are aligned with $(x,y)$. After that, the photons were transported through the patient geometry for dose calculations. Note that close to the boundary of the sampling area, some of the photons from those partially covered PSLs will be rejected by the jaws without going through particle transport.

S4. *Sample scatter photon sub-source PSLs.* Similarly, we first determine the number of particles using Eq. (3). The photon location and energy were uniformly distributed in its corresponding area and energy range. Once a photon location $(x,y)$ is determined, we first compute its radius $r = \sqrt{x^2 + y^2}$ and then an index $i$ in the scattered photon PSR model that it belongs to in order to obtain the corresponding parameters for the particle direction distribution model

$$p_{ps,i}(\alpha,\beta|r,E) = \sum_{k=1}^{K} G_{i,k}(\alpha,\beta) \tag{6}$$
$$= \sum_{k=1}^{K} \frac{c_{i,k}}{2\pi \times \sigma_{\alpha,i,k} \times \sigma_{\beta,i,k}} e^{-\frac{(\alpha-\mu_{i,k})^2}{2\sigma_{\alpha,i,k}^2} - \frac{\beta^2}{2\sigma_{\beta,i,k}^2}}.$$

Note that $i$ is a short notation for both energy and ring indices. Since this direction distribution for each scattered photon PSR is modeled with $K$ Gaussian components, we first sample the component index according to the relative weights $c_{i,k}$. After that, the two direction angles $\alpha$ and $\beta$, defined as $\alpha = \operatorname{atan}\left(\frac{u}{w}\right)$ and $\beta = \operatorname{atan}\left(\frac{v}{w}\right)$ in our source model, were sampled from the corresponding Gaussian component. Similarly, after particle direction being transformed, these scattered photons were then sent to a jaw-rejection module. Those photons surviving the jaw rejection were further transported in the patient geometry for dose calculations.

In our implementation, we have built a cumulative probability look-up table $P_{i,k} = \frac{\sum_{j=1}^{k} c_{i,j}}{\sum_{j=1}^{K} c_{i,j}}$ for $k = 1, \ldots K$ -1 to help sampling the index for the Gaussian components. With this table, the component index $k$ can be simply determined by generating a uniformly distributed number in [0,1] and search in this look-up table. Due to the sequential searching behavior, this step would lead to thread divergence, which becomes more severe with a large $K$. We set $K$ equal to 3 when building the direction distribution model to balance between sampling efficiency and model accuracy. The pre-calculated accumulative probability $P_{i,k}$ and the model parameters $\sigma_{\alpha,i,k}$, $\sigma_{\beta,i,k}$ and $\mu_{i,k}$ were transferred onto GPU at the code initialization stage and stored in the texture memory to take advantages of its fast access speed.

S5. Sample *electron sub-source PSLs*. We first determine the number of particles using Eq. (3). With the simple model of electron direction distribution

$$p_{e,i}(u,v|r,E) = p_{e,i}(\alpha,\beta|r,E) = \frac{1}{2\pi \times \sigma_{e\alpha} \times \sigma_{e\beta}} e^{-\frac{\alpha^2}{2\sigma_{e\alpha}^2} - \frac{\beta^2}{2\sigma_{e\beta}^2}}, \tag{7}$$

the sampling procedure is similar to that for the scattered photon PSLs. Hence, we do not repeat it here.

One issue during Step 3-5 is the order looping through all the PSLs. In GPU-based dose calculations, a user-specified number of threads $M$ are launched for a GPU kernel execution. Typically, one thread handles transport simulations of one particle. For instance, $M = 131072$ was set for our dose engine. Hence, it was not necessary to generate all the source particles from all the PSLs at the beginning and store them for later use. Doing so would require a lot of memory space. Instead, we iterated a GPU kernel of generating source particles and a kernel of particle transport. Each time $M$ source particles were processed. These $M$ particles may be from multiple PSLs, as the number $M$ is typically larger than the number of particles needed from a single PSL calculated in Eq. (3). To exhaust all the PSLs, we first looped over all the PSLs for different spatial locations and then over different energy bins. The purpose of this strategy was to keep particles simulated concurrently close in energy, which helps to avoid efficiency loss due to a few long GPU threads corresponding to particles with high energies.

*2.2 Automatic source commissioning*

*2.2.1 Basic principle*

For clinical application of our source model, it is important to commission the model to match an actual linac. The parameters in our model were originally derived from a reference phase-space file. We expect the model should be already very close to the actual linac beam, and hence the basic idea is to reweight each PSR in our analytical model to account for the difference between the reference phase-space file and the clinical beam. Specifically, a correction factor for each PSR was introduced. Doses of open fields for each PSR with its original weight derived from the reference phase-space file were pre-calculated in a water phantom, referred as "PSR dose". Because of the linearity of dose calculations, a dose distribution for a commissioned beam was simply a weighted sum of these pre-calculated PSR dose using the correction factors as weighting factors. These correction factors were automatically adjusted by a numerical algorithm that minimized the difference between a calculated dose in water and a measured dose.

As for the amount of measurement data required for commissioning, we propose to use depth dose at beam central axis and inline and cross-line beam profiles at several depths. These data for one large open field and one small field are needed. Compared to our previous approach (Townson *et al.*, 2013) that required data only at one large field size, more data are utilized here because of the more degrees of freedom in our beam model.

*2.2.2 Pre-calculating PSR dose in water*

Pre-calculating dose in water for each PSR is the first step for our commissioning approach. In our previous work under the PSL framework (Townson *et al.*, 2013), the MC dose calculation code was launched repeatedly, each time the source particles were only sampled from one specific PSL. The resulting 3D dose distribution was processed to extract the depth dose and lateral dose profiles at several depths necessary for commissioning. This strategy wasted a lot of time on initializing the MC code repeatedly and post-processing the 3D doses. To solve this problem, we made some modifications to our MC code, so that it is able to calculate the dose for all the PSRs simultaneously but store the required dose points for each PSR separately.

Specifically, we allocate space for a 2D dose array, in which each column is a list of data needed for beam commissioning for a PSR. Although there are a number of PSRs in our model, because full 3D dose is not needed anymore for each PSR, the required memory size still fits in the GPU memory. In addition, a 3D index array was created with the same size of the water phantom. Value zero for a voxel means that it is not required for commissioning. For those voxels needed for commissioning, their non-zero values are row indices in the 2D array that we allocated. This 3D array is loaded to GPU texture memory for fast access. During dose calculations, we sampled particles, as if doing a dose calculation for a given open field following the procedure described in Sec 2.1. However, each source particle carried an index regarding which PSR it is from. All secondary particles generated during transport simulation inherited the same index from their parent particle. Once a particle deposited dose to a voxel, we first checked if this voxel was needed for commissioning using the 3D index array. If so, the dose deposition was directed to the correct location in the 2D dose array determined by the particle PSR index and the row index in this voxel. This dose calculation was performed twice, yielding two dose matrices $A_1$ and $A_2$ corresponding to the large field and the small field cases needed for our model commissioning, respectively. Finally, we merge these two matrices to form a big matrix $A = [\begin{smallmatrix} A_1 \\ A_2 \end{smallmatrix}]$ to be used later in the commissioning process.

We would like to emphasize that the PSR dose were calculated in the presence of the original PSR weights derived from the reference phase-space file. Hence, the commissioning procedure determined a correction factor on top of them. Furthermore, because our commissioning model will involve data from one large and one small field sizes, dose calculations were performed in absolute dose sense, i.e. in a unit of cGy/MU, so that beam output can be correctly accounted.

*2.2.3 Commissioning model*

We mathematically formulate the automatic commissioning problem as:

$$x = min_{x \geq 0} F(x) = min_x \|Ax - b\|_2^2. \qquad (8)$$

Here, $x$ is a vector consisting of the correction factor for each photon PSR and each effective electron PSR unit (PSRs within same energy bin). They are nonnegative, since the meaning $x$ is a weighting factors for PSRs. Each column of the matrix $A$ corresponds to one PSR, containing depth dose and lateral profiles of two open fields calculated in the previous subsection. $b$ is a vector composed of measured data of a specific linac to be commissioned.

There are two reasons why we group the electron PSRs into different effective PSR units based on their energies for commissioning. First, the number of electrons for a photon beam is much smaller than that of photons (about 1%), and these electrons mainly contribute to build-up region, which is usually not a region of interests for photon beam dose calculations. Second, the limited amount of measurement data in shallow depths and its associated high uncertainty would inevitably impair the accuracy of the electron commissioning part.

*2.2.4 Optimization algorithm*

Gradient projection method was employed to solve the minimization problem with non-negativity constraints in Eq. (8). In this iterative method, a new solution $x^{k+1}$ was computed via by $x^{k+1} = P\left(x^k - s_k \times \nabla F(x^k)\right)$, where $k$ was the index for iteration step. $\nabla F(x^k)$ denoted gradient of the objective function $F(x)$, and $s_k$ was step size. $P(\cdot)$ denoted the projection operator to project a solution onto a feasible set. Here, for the non-negativity constraint, $P(y) = \max(0, y)$. Armijo's rule was adopted in a line search to determine the step size $s_k$ in order to balance the sufficient degree of accuracy and the convergence of the overall algorithm (Bazaraa *et al.*, 2013). If the relative change of the objective function between two successive iterations was smaller than a stopping tolerance $\delta$, the iteration will be stopped.

*2.3 Materials*

Our PSR-based analytical source model was originally constructed to represent a phase-space file of a Varian (Varian Medical System, Palo Alto, CA) TrueBeam 6MV beam (Capote, 2007). The resolution of the energy bins for our PSR model is 0.636 MeV (one tenth of the maximal particle energy in this reference phase-space file). The resolution of the PSR rings is 0.4 cm, which is the same to the resolutions in the $x$ and $y$ dimension when it comes to the PSL representation used in dose calculations.

The dose engine we used in this paper is our MC dose calculation package newly developed under an OpenCL environment (Tian *et al.*, 2015b), a cross-platform package and can be run on CPU, GPUs from different vendors, and even heterogeneous platforms (Khronos OpenCL Working Group, 2013). Transport cutoff energies for electrons and photons in MC simulation were 200 KeV and 50 KeV, respectively.

The first purpose of this paper was to demonstrate the efficiency gain of using our analytical source model together with the sampling procedure in GPU-based MC dose calculations. Therefore, the dose calculations for different open fields in a water phantom and one prostate patient IMRT case and one Head-and-neck (H&N) IMRT case were performed using our MC dose engine with our analytical source model. For comparison purpose, dose calculations using the previous PSL-file source model were also conducted. The water phantom had a size of about 60×60×60 cm$^3$ and a resolution of 0.25×0.25×0.2 cm$^3$. Several open fields with sizes 2×2 cm$^2$, 5×5 cm$^2$, 10×10 cm$^2$, 20×20 cm$^2$ and 40×40 cm$^2$ were used, with source-to-surface distance (SSD) set to 100 cm. For the prostate patience case, there were 7 coplanar beams in the IMRT treatment plan. The voxel size was 0.195×0.195×0.25 cm$^3$. The H&N patient case had 5 coplanar beams at 0° couch angle and 1 non-coplanar beam at 90° couch angle. The voxel size was 0.137×0.137×0.125cm$^3$. The source-to-axis distances (SAD) for both cases were 100 cm.

The efficacy of our automatic commissioning method was also demonstrated. The same water phantom mentioned above was used in pre-calculating the dose sets of each PSR sub-source for commissioning. The large field size used for commissioning was 40×40 cm$^2$ and the small field size was 5×5 cm$^2$. For each field size, depth dose at the beam central axis and inline and crossline lateral dose profiles at depth 1.5 cm, 10 cm and 20 cm were used. For the measurement, data was acquired from a clinical Varian TrueBeam machine in our institution. After commissioning, depth dose, lateral profiles, and output factors for a set of fields ranging from 2×2 cm$^2$ to 40×40 cm$^2$ were computed using our MC dose engine, which were then compared with measurements for validation purpose.

## 3. Results

*3.1 Efficiency test of our PSR-based analytical source model*

The computational efficiency of our PSR-based analytical source model for GPU-based MC dose calculation was investigated and compared with that using the PSL-files as source model. Here we didn't compare with the efficiency when a phase-space file was directly used for dose calculations, since efficiency of that approach was already demonstrated to be much inferior to the PSL approach (Townson *et al.*, 2013). The total computation time of MC dose calculation $T_{total}$ for different cases with 1 billion source particles were recorded and listed in Table 1. To better understand the efficiency gain of our PSR-based analytical source model for MC dose calculation, we broke down the total computational time further and list the particle transport simulation time $T_{sim}$ and the times for some steps related to the source model. Specifically, for the PSL source model, time spent on initializing the model $T_{src\_init}$, time of reading the particles from these PSL files $T_{par\_rd}$, that of transferring these particle data from CPU to GPU $T_{par\_cp}$, and that of translating and rotating these particles according to beam setup $T_{par\_tr}$ were recorded. For our analytical source model, we record the time for its initialization, and the cumulative time $T_{par\_gen\_tr}$ to generate the particles on GPU and translate/rotate these particles for transport.

It was observed from Table 1 that, when using PSL files as source model, a lot of time was actually spent on reading particles from PSL files stored on the computer hard drive, ranging from 71.85 to 148.43 seconds depending on different jaw openings. This data reading time accounted for 38.4%~75.7% of the total computation time. Besides, it also took a few seconds to transfer all the source particles from CPU to GPU. Initialization of this source model and translation/rotation of the particles were finished in milliseconds and the time could be ignored. In contrast, for our PSR-based analytical source model, the particles were generated on the fly on GPU, which avoided long data reading times from the hard drive and data transferring from CPU to GPU. The time needed to generate 1 billion particles and translate/rotate them for transport simulation was about 4 seconds, accounting for only 3.6%~8.6% of the total computation time. The initialization of our source model was also done within a few milliseconds. The efficiency was improved by our analytical source model with a factor of 1.70 ~ 4.41 compared to that of using PSL files as a source model. This comparison has clearly demonstrated that the utilization of our PSR-based analytical source model greatly increased the overall efficiency of MC dose calculation on GPU platform compared to directly using the PSL source model.

**Table 1.** Efficiency test results of using our PSR-based analytical source model verses using PSL files as source model. $T_{total}$ denotes the total computation time of MC dose calculation using our new dose engine; $T_{sim}$ denotes the time spent on particle transport simulation; $T_{src\_init}$ denotes the time spent on initializing the source model (either PSL files or our analytical model); $T_{par\_rd}$, $T_{par\_cp}$ and $T_{par\_tr}$ denote the time spent on reading particles from the PSL files, transfering these particle data from CPU to GPU and translating and rotating the particles according to beam setup, respectively, when using the PSL files as source model; $T_{par\_gen\_tr}$ denote the time for particle generation, translation and rotation when using our analytical model.

| Case | | Using PSL files as source model | | | | | PSR-based analytical source model | | | |
|---|---|---|---|---|---|---|---|---|---|---|
| | | $T_{total}$ (s) | $T_{sim}$ (s) | $T_{src\_init}$ (s) | $T_{par\_rd}$ (s) | $T_{par\_cp}$ (s) | $T_{par\_tr}$ (s) | $T_{total}$ (s) | $T_{sim}$ (s) | $T_{src\_init}$ (s) | $T_{par\_gen\_tr}$ (s) |
| Water phantom | 40×40 | 189.60 | 104.38 | 0.02 | 72.83 | 3.85 | 0.15 | 111.81 | 105.62 | 0.01 | 4.00 |
| | 20×20 | 144.68 | 62.67 | 0.03 | 71.85 | 1.93 | 0.04 | 68.32 | 62.39 | 0.01 | 3.69 |
| | 10×10 | 185.87 | 63.71 | 0.02 | 111.93 | 1.68 | 0.07 | 69.15 | 62.96 | 0.00 | 3.92 |
| | 5×5 | 209.14 | 62.99 | 0.02 | 135.93 | 1.69 | 0.05 | 70.31 | 64.07 | 0.00 | 4.02 |
| | 2×2 | 183.90 | 37.03 | 0.02 | 138.57 | 0.71 | 0.01 | 42.38 | 36.52 | 0.01 | 3.64 |
| Prostate case | | 196.08 | 38.11 | 0.02 | 148.43 | 1.48 | 0.05 | 44.50 | 38.88 | 0.01 | 3.73 |
| H&N case | | 181.15 | 40.86 | 0.03 | 130.88 | 1.24 | 0.03 | 45.76 | 40.29 | 0.01 | 3.84 |

*3.2 Efficacy test of our source commissioning method*

We have previously demonstrated that the PSR-based analytical source model is capable of representing the reference phase-space file (Tian *et al.*, 2015a). Here it is our purpose to show that, with the automatic commissioning approach, we can further tune the beam model to represent an actual linac beam. As such, we first pre-computed the PSR doses under two open fields, i.e. 40×40 $cm^2$ and 5×5 $cm^2$. Using measurements at these two field sizes, we ran our automatic commissioning procedure. Because of the small problem size, the optimization problems were solved in ~20 seconds using Matlab (MathWorks, Natick, MA), yielding correction factors for each PSR.

After the beam model was commissioned, the dose distributions were calculated for five open fields (40×40 $cm^2$, 20×20 $cm^2$, 10×10 $cm^2$, 5×5 $cm^2$, 2×2 $cm^2$) with SSD 100 cm. The calculated depth dose curves and the inline and crossline lateral dose profiles at three depths (1.5 cm, 10 cm, 20 cm) were compared with the measurement data, as well as with the dose calculated with the reference phase-space file. The comparison results were shown in Figure 5 and Figure 6, respectively. From Figure 5, obvious discrepancies of depth dose curves were observed between the reference phase-space file and the measurements. The difference was particularly obvious for the build-up region of large fields, which implied that the energy spectrum of the contaminant electrons in the reference phase-space may be difference from that of the linac machine. In contrast, our commissioned results showed a better match with the measurement data in the build-up region. Although we adopted a simplified commissioning model for electrons to only adjust the energy spectrum of electrons, our experimental results demonstrated that this simplified model was able to provide clinically acceptable accuracy. From Figure 6, large differences of the profile shape between the reference phase-space file and the measurement were found for these two large fields in both inline and crossline directions, as indicated by the arrows. Obvious dose discrepancies were also found at the outer beam region of the inline dose profiles for the 20×20 $cm^2$ open field, as indicated by the arrows. By reweighting PSRs in our source model, our commissioning approach achieved a better match of dose profiles with the measurement.

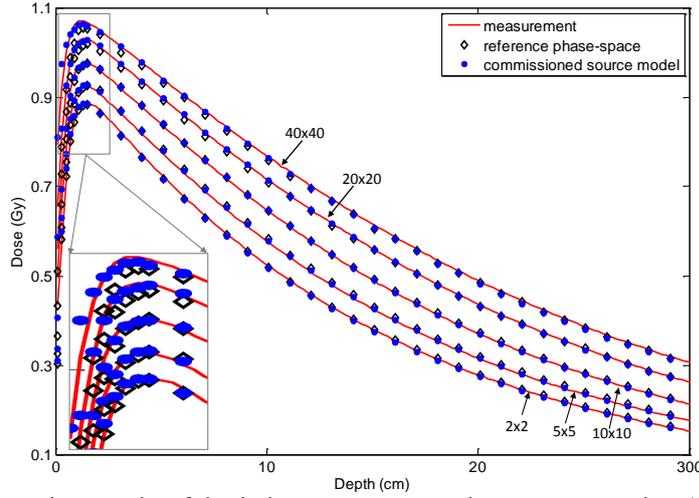

**Figure 5.** The comparison results of depth dose curves among the measurement data (solid line), the data calculated with the reference phase-space file (open diamond) and those calculated with our commissioned analytical source model (dots). Five open fields (40×40 $cm^2$, 20×20 $cm^2$, 10×10 $cm^2$, 5×5 $cm^2$, 2×2 $cm^2$) with SSD 100 cm are compared here.

These results were further quantitatively compared using different comparison metrics for different regions as suggested by AAPM task group 53 (Fraass *et al.*, 1998). Specifically, the root-mean-square (RMS) difference and the maximum difference were calculated for the dose regions with low gradient, such as the depth dose after build-up and the lateral dose profiles at inner beam and outer beam regions. The calculations of these two metrics are as follows:

$$RMS(\%) = \frac{1}{D_{max}^m}\sqrt{\frac{1}{N}\sum_{i=1}^{N}(D_i^c - D_i^m)^2}, \quad (10)$$

$$Max(\%) = \frac{1}{D_{max}^m}max_i|D_i^c - D_i^m|. \quad (11)$$

Here, $D_i^c$ denotes the dose value of the $i_{th}$ comparison dose point for the dose calculated with either our analytical source model or the reference phase-space file. $D_i^m$ denotes the corresponding measurement data, regarded as the ground truth for comparison. $D_{max}^m$ denotes measured depth dose at $d_{max}$. For high-gradient regions, such as the depth dose at build-up region and lateral dose profiles at penumbra region, we employed distance-to-agreement (DTA) for evaluation. DTA at a spatial location $x$ is defined to be the minimum distance $s = |x - y|$ such that $D^c(y) = D^m(x)$.

These quantitative comparison results were presented in Table 2. For depth dose curves, our commissioned source model reduced the average DTA of the build-up region from 0.039~0.280 cm to 0.039~0.117 cm and improved the maximum DTA from 0.086~0.472 cm to 0.097~0.365 cm. For the depth dose after build-up, RMS was reduced from 0.323%~0.670% to 0.207%~0.484% and the maximum difference from 0.639%~1.586% to 0.582%~0.930%. For the smallest field 2×2 $cm^2$, the commissioned beam deviates more from the ground truth than the reference phase-space file, the accuracy is still clinical acceptable with maximum DTA less than 0.1 cm, RMS less than 0.5% and the maximum difference less than 1%. In addition, the measurements for this small field size are very challenging and relatively large uncertainty may exist.

For the comparison results of inline dose profiles shown in Table 3, our commissioned source model achieved a better match with the measurement data, improving RMS and maximum difference at inner beam region from 0.307%~2.007% to 0.061%~0.779% and from 0.468%~3.217% to 0.148% ~3.213% respectively, and those at outer beam region from 0.196%~1.251% to 0.099%~0.511% and from 0.270%~1.950% to 0.153%~0.754% respectively. For the penumbra region of inline dose profiles, since the reference phase-space file already achieved a good match with the measurement data, our commissioned source model didn't improve too much. For the crossline dose profiles, we found the similar level of improvements, which we do not present in detail due to space limitations.

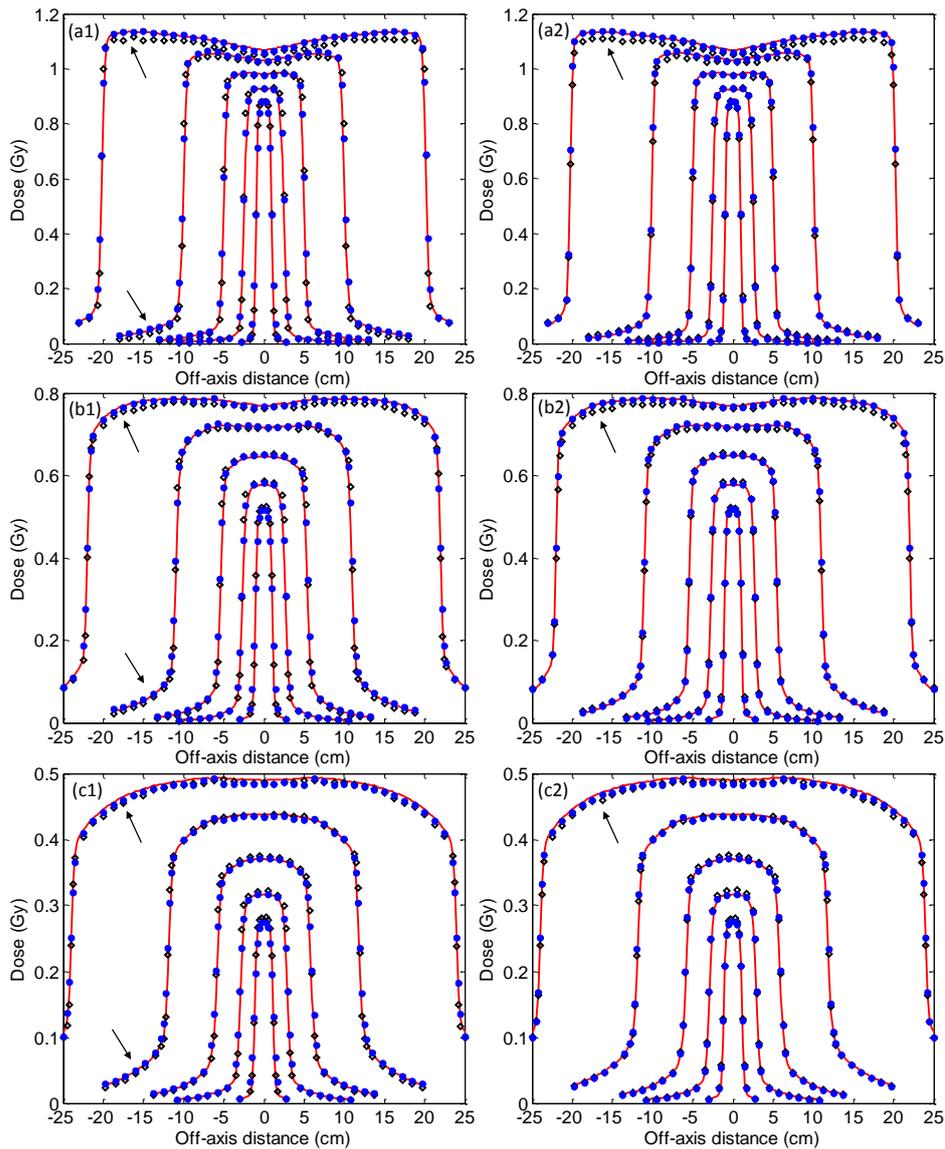

**Figure 6.** The comparison results of the inline and crossline lateral dose profiles among the measurement data (solid line), the data calculated with the reference phase-space file (dimond) and those calculated with our commissioned analytical source model (dots). Five open fields (40×40 $cm^2$, 20×20 $cm^2$, 10×10 $cm^2$, 5×5 $cm^2$, 2×2 $cm^2$) with SSD 100 cm are compared here. (a1~c1) show the inline dose profiles of those open fields at depth 1.5 cm, 10 cm and 20 cm, respectively. (a2~c2) show the corresponding crossline dose profiles. Same legend as in Figure 8 is used.

**Table 2.** Quantitative comparison results of depth dose curves between the dose calculated with the reference phase-space file directly $D_{ref}$ and the measurement data and between the dose calculated with our commissioned source model $D_{com}$ and the measurement data.

| Field size ($cm^2$) | Build-up region | | | | Region after build-up | | | |
|---|---|---|---|---|---|---|---|---|
| | Average DTA(cm) | | Maximum DTA(cm) | | RMS(%) | | Max(%) | |
| | $D_{ref}$ | $D_{com}$ | $D_{ref}$ | $D_{com}$ | $D_{ref}$ | $D_{com}$ | $D_{ref}$ | $D_{com}$ |
| 40×40 | 0.278 | 0.117 | 0.472 | 0.365 | 0.655 | 0.478 | 1.554 | 0.900 |
| 20×20 | 0.178 | 0.067 | 0.363 | 0.204 | 0.434 | 0.194 | 1.586 | 0.582 |
| 10×10 | 0.100 | 0.064 | 0.132 | 0.109 | 0.323 | 0.207 | 0.639 | 0.593 |
| 5×5 | 0.068 | 0.038 | 0.120 | 0.109 | 0.670 | 0.366 | 1.131 | 0.773 |
| 2×2 | 0.039 | 0.039 | 0.086 | 0.097 | 0.425 | 0.484 | 0.869 | 0.930 |

**Table 3.** Quantitative comparison results of inline dose profiles between the dose calculated with the reference phase-space file directly $D_{ref}$ and the measurement data and between the dose calculated with our commissioned source model $D_{com}$ and the measurement data.

| Field size ($cm^2$) | Depth (cm) | | Penumbra | | Inner beam | | Outer beam | |
|---|---|---|---|---|---|---|---|---|
| | | | Average DTA(cm) | Maximum DTA(cm) | RMS(%) | Max(%) | RMS(%) | Max(%) |
| 40×40 | 1.5 | $D_{ref}$ | 0.077 | 0.120 | 2.007 | 2.881 | 1.030 | 1.618 |
| | | $D_{com}$ | 0.068 | 0.108 | 0.428 | 1.231 | 0.304 | 0.487 |
| | 10 | $D_{ref}$ | 0.044 | 0.132 | 1.108 | 1.949 | 0.634 | 0.752 |
| | | $D_{com}$ | 0.033 | 0.068 | 0.315 | 0.801 | 0.321 | 0.403 |
| | 20 | $D_{ref}$ | 0.037 | 0.080 | 0.657 | 1.297 | 0.973 | 1.028 |
| | | $D_{com}$ | 0.039 | 0.064 | 0.548 | 1.033 | 0.511 | 0.661 |
| 20×20 | 1.5 | $D_{ref}$ | 0.078 | 0.121 | 1.161 | 3.217 | 1.251 | 1.481 |
| | | $D_{com}$ | 0.071 | 0.112 | 0.779 | 3.213 | 0.347 | 0.525 |
| | 10 | $D_{ref}$ | 0.095 | 0.169 | 0.448 | 1.223 | 0.629 | 0.697 |
| | | $D_{com}$ | 0.083 | 0.117 | 0.433 | 0.883 | 0.324 | 0.393 |
| | 20 | $D_{ref}$ | 0.099 | 0.183 | 0.307 | 0.641 | 0.427 | 0.518 |
| | | $D_{com}$ | 0.087 | 0.129 | 0.314 | 0.562 | 0.099 | 0.153 |
| 10×10 | 1.5 | $D_{ref}$ | 0.078 | 0.144 | 0.598 | 1.086 | 0.858 | 1.581 |
| | | $D_{com}$ | 0.075 | 0.110 | 0.436 | 1.085 | 0.281 | 0.542 |
| | 10 | $D_{ref}$ | 0.101 | 0.223 | 0.706 | 1.277 | 0.410 | 0.645 |
| | | $D_{com}$ | 0.094 | 0.120 | 0.464 | 0.902 | 0.276 | 0.754 |
| | 20 | $D_{ref}$ | 0.100 | 0.211 | 0.515 | 1.078 | 0.245 | 0.579 |
| | | $D_{com}$ | 0.093 | 0.129 | 0.061 | 0.148 | 0.141 | 0.605 |
| 5×5 | 1.5 | $D_{ref}$ | 0.076 | 0.132 | 0.599 | 1.380 | 0.601 | 1.144 |
| | | $D_{com}$ | 0.073 | 0.103 | 0.552 | 1.429 | 0.284 | 0.697 |
| | 10 | $D_{ref}$ | 0.084 | 0.136 | 1.187 | 2.267 | 0.302 | 0.464 |
| | | $D_{com}$ | 0.078 | 0.107 | 0.673 | 1.364 | 0.118 | 0.358 |
| | 20 | $D_{ref}$ | 0.086 | 0.137 | 0.700 | 1.462 | 0.196 | 0.270 |
| | | $D_{com}$ | 0.080 | 0.112 | 0.134 | 0.296 | 0.123 | 0.308 |
| 2×2 | 1.5 | $D_{ref}$ | 0.069 | 0.108 | 0.761 | 0.915 | 0.200 | 0.278 |
| | | $D_{com}$ | 0.073 | 0.116 | 0.649 | 0.708 | 0.214 | 0.302 |
| | 10 | $D_{ref}$ | 0.083 | 0.282 | 0.579 | 1.140 | 0.708 | 1.684 |
| | | $D_{com}$ | 0.100 | 0.305 | 0.530 | 1.071 | 0.164 | 0.323 |
| | 20 | $D_{ref}$ | 0.105 | 0.313 | 0.399 | 0.468 | 0.836 | 1.950 |
| | | $D_{com}$ | 0.120 | 0.341 | 0.325 | 0.434 | 0.171 | 0.388 |

The output factors of our commissioned source model were also calculated and compared with those of the measurement data and the reference phase-space file, as shown in Figure 7. The relative differences compared with the measurement data was reduced from within 0.668% to within 0.259%.

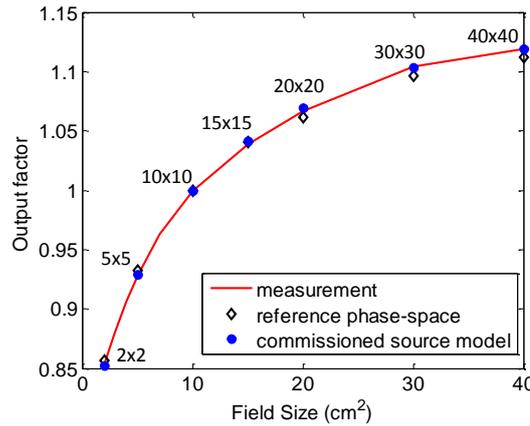

**Figure 7.** Output factors of the measurement data (solid line), the reference phase-space file (open diamond) and those of our commissioned analytical PSL-based source model (dots).

## 4. Discussion and Conclusions

In this paper, we have presented methods to efficiently utilize our PSR-based analytical source model that we have recently built for photon beams in GPU-based MC dose calculation. As such, our PSR model was first converted to its PSL representation. Then a GPU-friendly source sampling strategy was employed, which separately sampled different types of sub-sources and iterated the kernel of source sampling and kernel of particle transport during dose calculations. This strategy ensured that the particles being sampled and transported simultaneously were always of the same type and close in energy, which could alleviate GPU thread divergence during both the source sampling and particle transport simulation. The second purpose of this paper was to present an automatic commissioning approach for our analytical source model to automatically and flexibly adjust the source model in order to achieve a better match with realistic measurement data acquired from a clinical linac machine. In our method, weighting factors were introduced to adjust relative weight of each sub-source. Determining these factors was realized by solving a quadratic minimization problem with a non-negativity constraint, which could be achieved within about half minute. We have tested the efficiency gain of our analytical source model over the source using PSL files. It was found that the efficiency was improved by our analytical source model with a factor of 1.70 ~ 4.41 in phantom and real patient cases, mainly due to the avoidance of long data reading and CPU-to-GPU data transferring. The automatic commissioning problem can be solved in ~20 sec. Its efficacy was tested by comparing the doses computed using the commissioned model, the uncommissioned one, with measurements in different open fields in a water phantom under a clinical Varian Truebeam 6MV beam. For the depth dose curves, the average distance-to-agreement (DTA) was improved from 0.04~0.28 cm to 0.04~0.12 cm for build-up region and the root-mean-square (RMS) dose difference after build-up region was reduced from 0.32%~0.67% to 0.21%~0.48%. For the lateral dose profiles, RMS difference was reduced from 0.31%~2.0% to 0.06%~0.78% at inner beam region and from 0.20%~1.25% to 0.10%~0.51% at outer beam region.

There are several major improvements of this new commissioning method for our PSR-based analytical source model over the one we developed previously for the source model using PSL files (Townson *et al.*, 2013). First, there was no regularization terms needed, as the beam symmetric property has already been imposed by our PSR-based model. This made our commissioning model simpler and easier to solve. Second, separation of primary and scattered photons in our PSR beam model allowed fine-tuning of the ratio between them in commissioning. Third, instead of considering all the electrons as a whole unit for commissioning, we regarded the electron PSRs within a same energy bin as one effective electron PSR unit, so that we can have more degrees of freedom to commission electrons. Fourth, as opposed to sequentially commissioning photon and electron components, they were considered simultaneously in an optimization problem. Last but not least, the MC code was modified to pre-calculate the dose of all the PSR sub-sources simultaneously but store them separately. This strategy saved time on code initialization, data transfer from CPU to GPU and data post-processing.

One interesting results found in our previous automatic model commissioning study for the PSL-based source model was that a PSL data set generated for a linac from a particular vendor can be used to commission linacs from other vendors. This was probably ascribed to similar design above the jaws among different linacs. We have not tested this issue for our new source model. However, we expect this fact will still hold. Besides, the concept of our source model and commissioning method should be also applicable to the flattening filter free (FFF) beam. Eliminating the flattening filter may even simplify our source model by avoiding the scattered photon term.

Although our commissioning method was proposed for our PSR-based analytical source model, the commissioning result was not restricted to dose calculations using the model. In fact, a new phase-space file could be generated based on our corrected PSR weighting factors and adopted by other MC dose calculation tools that directly use a phase-space file as a beam model.

**Acknowledgements**

We would also like to thank ….